\documentclass[preprint]{aastex}
\usepackage{amsmath}
\usepackage{graphicx} 
\usepackage{subfigure}
\usepackage{float}

\begin{document}

\title{Point Source Polarimetry with the Gemini Planet Imager: Sensitivity Characterization with T5.5 Dwarf Companion HD 19467 B}

\author{Rebecca Jensen-Clem\altaffilmark{1}, Max Millar-Blanchaer\altaffilmark{2}, Dimitri Mawet\altaffilmark{1}, James R. Graham\altaffilmark{3}, J. Kent Wallace\altaffilmark{4}, Bruce Macintosh\altaffilmark{5},  Sasha Hinkley\altaffilmark{6},  Sloane J. Wiktorowicz\altaffilmark{7}, Marshall D. Perrin\altaffilmark{8}, Mark S. Marley\altaffilmark{9},  Michael P. Fitzgerald\altaffilmark{10}, Rebecca Oppenheimer\altaffilmark{11}, S. Mark Ammons\altaffilmark{12}, Fredrik T. Rantakyr\"o\altaffilmark{13}, Franck Marchis\altaffilmark{14}}

\altaffiltext{1}{Department of Astrophysics, California Institute of Technology, 1200 E. California Blvd., Pasadena, CA 91101, USA}
\altaffiltext{2}{Department of Astronomy \& Astrophysics, University of Toronto, Toronto, ON, Canada}
\altaffiltext{3}{Department of Astronomy, UC Berkeley, Berkeley, CA 94720, USA}
\altaffiltext{4}{Jet Propulsion Laboratory, California Institute of Technology, 4800 Oak Grove Drive, Pasadena, CA, USA 91109}
\altaffiltext{5}{Department of Physics \& Kavli Institute for Particle Astrophysics and Cosmology, Stanford University, Palo Alto, CA, USA}
\altaffiltext{6}{University of Exeter, Physics Department, Stocker Road, Exeter, EX4 4QL, UK}
\altaffiltext{7}{Department of Astronomy, UC Santa Cruz, 1156 High Street, Santa Cruz, CA 95064, USA}
\altaffiltext{8}{Space Telescope Science Institute, 3700 San Martin Drive, Baltimore, MD 21218, USA}
\altaffiltext{9}{NASA Ames Research Center, MS-245-3, Moffett Field, CA 94035, USA}
\altaffiltext{10}{Department of Physics and Astronomy, UCLA, Los Angeles, CA 90095, USA}
\altaffiltext{11}{American Museum of Natural History, New York, NY 10024, USA}
\altaffiltext{12}{Lawrence Livermore National Laboratory, Livermore, CA 94551, USA}
\altaffiltext{13}{Gemini Observatory, Casilla 603, La Serena, Chile}
\altaffiltext{14}{SETI Institute, Carl Sagan Center, 189 Bernardo Avenue, Mountain View, CA 94043, USA}

\begin{abstract}

Detecting polarized light from self-luminous exoplanets has the potential to provide key information about rotation, surface gravity, cloud grain size, and cloud coverage. While field brown dwarfs with detected polarized emission are common, no exoplanet or substellar companion has yet been detected in polarized light. With the advent of high contrast imaging spectro-polarimeters such as GPI and SPHERE, such a detection may now be possible with careful treatment of instrumental polarization. In this paper, we present 28 minutes of $H$-band GPI polarimetric observations of the benchmark T5.5 companion HD 19467 B. We detect no polarization signal from the target, and place an upper limit on the degree of linear polarization of $p_{\text{CL}99.73\%} \leq 2.4\%$. We discuss our results in the context of T dwarf cloud models and photometric variability.

\end{abstract}

\section{Introduction}

Recent imaging spectroscopy of directly imaged planets such as $\beta$ Pic b, HR 8799 bcd, and 51 Eri b \citep{Chilcote2015, Ingraham2014, Macintosh2015} has demonstrated that this technique will soon become routine.  At the same time, these early studies have underscored the challenges of fitting atmospheric models based on field brown dwarf spectra to exoplanets with comparatively low masses and surface gravities \citep{Barman2011, Marley2012}. For instance, the HR 8799 planets' red colors are consistent only with models that include dusty, patchy clouds and non-equilibrium chemistry, but these scenarios require radii that are inconsistent with the predictions of evolutionary models \citep{Marois2008b, Bowler2010, Ingraham2014}.  The spectra of $\beta$ Pic b gathered to date also require a cloudy atmosphere, but even the best fitting models contain large ($5\%-10\%$) systematic offsets from the data \citep{Chilcote2015}. These and other studies indicate that clouds are common in planetary atmospheres, but current atmospheric models require a more detailed treatment of cloud physics (e.g. dust grain size distributions, grain chemistry, and large scale opacity holes, or ``patchy" cloud regions). 

Polarimetry is well-suited to the problem of analyzing planetary atmospheres. As early as 1929, Bernard Lyot published reflected light polarimetry of Venus as a means of probing the planet's cloud composition \citep{Lyot1929}. In the modern era of exoplanet astronomy, at least 10 years of theoretical work have been devoted to the polarization of \emph{reflected starlight} by a close-in directly imaged exoplanet \citep{Schmid2006, Stam2008, Buenzli2009}. In this paper, we address a separate regime of exoplanet polarimetry: scattering by grains in the atmospheres of \emph{cloudy, self-luminous} exoplanets,  which induces polarization of \emph{thermally emitted radiation} in the near infrared \citep{Sengupta2009, Marley2011, deKok2011}.  While Rayleigh scattering in the atmospheres of cloud-free exoplanets can polarize visible wavelength radiation, we concentrate on the case of $\sim$micron sized dust grains  that polarize infrared radiation, where young, self-luminous exoplanets emit the bulk of their radiation.  

A perfectly spherical, uniformly cloudy planet, however, would have zero disk-integrated polarization. In order to produce a non-zero polarization signature when the planet is viewed as a point source, the clouds must be nonuniformly distributed (e.g. patchy or banded) and/or the body must be oblate \citep{Basri2000, Sengupta2001}. Both of these characteristics have been invoked to explain the observed polarization of brown dwarfs in the field. Time-varying polarimetric signals indicate evolving or rotating nonuniform cloud distributions, while comparisons of fast and slow brown dwarf rotators show that oblate bodies are more likely to produce observable polarimetric signals than spherical bodies \citep{Menard2002, Osorio2005, Goldman2009, Tata2009, Osorio2011, MilesPaez2013}. For example, \citet{MilesPaez2013} found that $40 \pm 15 \%$ of rapidly rotating ($v\,$sin$\,i\geq 30\,$km$\,$s$^{-1}$) M7-T2 dwarfs are linearly polarized in the $Z-$ and $J-$bands, with linear degrees of polarization of $0.4\%-0.8\%$. The fastest rotators ($v\,$sin$\,i> 60\,$km$\,$s$^{-1}$) were more frequently polarized, and had larger detected polarizations compared to moderate rotators ($v\,$sin$\,i= 30-60\,$km$\,$s$^{-1}$). Furthermore, polarimetry in different broad wavelength bands has provided a diagnostic of grain sizes \citep{Osorio2005}.

These field observations set the stage for exoplanet and brown dwarf companion polarimetry. Because models and field dwarf observations agree that the cloud-induced degree of linear polarization is generally $p\leq 1\%$, the observational challenge is to reach the companion-to-star contrast ratio and absolute polarimetric accuracy needed to detect such small signals. SPHERE \citep{Beuzit2008} and GPI \citep{Macintosh2008, Macintosh2014} are the first spectro-polarimeters capable of achieving contrasts $<10^{-5}$ at separations $<1$" and an absolute accuracy in the degree of linear polarization of $p<0.1\%$ in the near infrared \citep{Wiktorowicz2014}. These instruments are poised to make the first detections of polarized light from substellar companions, providing a powerful new tool for atmospheric characterization. 

In order to assess GPI's accuracy in linear polarization for point sources, we observed the benchmark T dwarf HD 19467 B for 28 minutes of integration time on February 1st 2015. HD 19467 B is unique among substellar companions in that it is the only T dwarf with a solar-analog primary to be detected both as a long term trend in RV measurements and by direct imaging \citep{Crepp2012}.  Recently, \citet{Crepp2015} determined a spectral type of T$5.5 \pm 1$ and an effective temperature of $T_{\text{eff}} = 978^{+20}_{-43}\,$K (see Table \ref{tbl:params} below) using the Project 1640 integral field spectrograph at Palomar Observatory \citep{Hinkley2011}. While mid T dwarfs are generally thought to be cloudless, HD 19467 B was the only brown dwarf companion or exoplanet meeting GPI's observability requirements ($I<9$ parent star and planet-star separation $< 2.0"$) at the time of the pathfinder experiment.

We describe our observational methods and Stokes parameter extraction in Section \ref{sec:obs}. In Section \ref{sec:pol} we measure the degree of linear polarization at the location of the companion and empirically estimate GPI's polarimetric point source sensitivity. In Section \ref{sec:discussion}, we place our results in the context of field T dwarf observations and discuss our upcoming GPI campaign for exoplanet polarimetry.

\begin{table}[h]
\centering
\caption{HD 19467 system properties, after \citet{Crepp2014, Crepp2015} }
\begin{tabular}{ p{6cm} p{4cm}}
\hline \hline
\multicolumn{2}{c}{HD 19467 A Properties} \\
\hline 
$J$  & $5.801 \pm 0.020$ \\
$H$ & $5.447 \pm 0.036$ \\
$K_{s}$ & $5.401 \pm 0.026$ \\
Mass [M$_{\odot}$] & $0.95 \pm 0.02$ \\
Radius [R$_{\odot}$] & $1.15 \pm 0.03$ \\
Luminosity [L$_{\odot}$] & $1.34 \pm 0.08$ \\
$T_{\text{eff}}$ [K]  & $5680 \pm 40$ \\
SpT & G3V \\
d [pc] & $ 30.86 \pm 0.60$ \\
Age, multiple techniques [Gyr] & 4.6 - 10 \\
$[\text{Fe/H}]$ & $-0.15 \pm 0.04$ \\
\hline \hline
\multicolumn{2}{c}{HD 19467 B Properties} \\
\hline 
$J$  & $ 17.61 \pm 0.11 $ \\
$H$ & $17.90 \pm 0.11 $ \\
$K_{s}$ & $ 17.97 \pm 0.09 $ \\
Mass [$M_{\text{Jup}}$] & $\geq 51.9^{+3.6}_{-4.3}$ \\
$T_{\text{eff}}$ [K] & $978^{+20}_{-43}$ \\
SpT & T$5.5 \pm 1$ \\
Separation [AU, as] & $51.1 \pm 1.0$, 1.65" \\
\hline
\multicolumn{2}{l}{\textbf{References.} \citet{Crepp2014, Crepp2015}.} 

\end{tabular}
\label{tbl:params}
\end{table}

\section{Observations and Data Reduction}
\label{sec:obs}

We briefly summarize GPI's polarimetric mode here; for details, see \citet{Macintosh2014}, \citet{Wiktorowicz2014}, and \citet{Perrin2015}. In polarimetry mode, a Wollaston prism replaces the integral field spectrograph's dispersing prism, and an achromatic half waveplate is inserted between the coronagraph's focal plane and Lyot mask. In an observing sequence, the waveplate is rotated after each exposure in steps of $22.5^{\circ}$. GPI operates in angular differential imaging (ADI) mode, allowing the sky to rotate with respect to the telescope pupil. The Stokes datacube $[x, y, (\mathrm{I,Q,U,V)}]$ describing the astronomical polarization is extracted from the raw data by inverting the Mueller matrix whose elements represent the polarization induced by the instrument and sky rotation. A detailed description of the Stokes cube extraction can be found in \citet{Perrin2015}. 

The observing sequence consisted of twenty-eight  60s exposures in the $H$-band, where each sequence of four exposures cycled through the waveplate angles $0.0^{\circ}$, $22.5^{\circ}$, $45.0^{\circ}$, and $67.5^{\circ}$. The exposures were taken from Modified Julian Day 57054.0536728 to 57054.0773418 and the airmass ranged from 1.21 to 1.32. The total field rotation was $4^{\circ}$. The observations were taken under Gemini program number GS-2014B-Q-503. 

In this study, the Stokes datacube was constructed from the raw data using a modified version of the publicly available GPI pipeline v1.2.1 \citep{Maire2010, Perrin2014}. Modifications, which will be released in a future version of the pipeline, included improved flat fielding and instrumental polarization subtraction. The latter was achieved by measuring the fractional polarization inside the coronagraphic mask, which was assumed to contain only diffracted starlight affected by instrumental polarization, on each polarization difference cube. The instrumental polarization was then subtracted from the image as a whole, by multiplying the measured fractional polarization by the total intensity at each spatial location. This method was introduced by \citet{MillarBlanchaer2015}, who found similar instrumental polarization levels as those measured using the same dataset by \citet{Wiktorowicz2014}. The method used in \citet{Wiktorowicz2014} to estimate the instrumental polarization measures the fractional polarization in each frame using aperture photometry on the occulted PSF, and fits an instrumental polarization model that leverages the field rotation within their dataset to distinguish between an astrophysical source and the instrumental polarization. The method used here and in \citet{MillarBlanchaer2015} has the advantage of operating on each polarization datacube individually and therefore, this method doesnÕt rely on sky rotation. While promising for point source polarimetry data, techniques such as LOCI or PCA that take greater advantage of the sky rotation to reduce instrumental polarization are beyond the scope of this paper. Section \ref{sec:pol} briefly describes how our results could be improved by these techniques. 

Spurious pixels in the Stokes cube were replaced with values obtained by interpolation over pixels with counts greater than three standard deviations from the mean, and the detectorÕs gain of 3.04e-/ADU was applied to convert the raw counts to electrons (this gain is reported in the FITS headers of all GPI data cubes). The final reduced Stokes I, Q, and U images are shown in Figure 1. The four bright areas in the corners of the Stokes I frame (Figure 1a) are satellite spots used for photometric and astrometric calibration in GPI's imaging spectroscopy mode. 


HD 19467 B is readily visible on the righthand side of the Stokes I frame, but is not apparent in the Stokes Q or U frames. In the next section, we will calculate the companion's signal to noise ratio in each Stokes frame, and comment on the degree of linear polarization at the companion's location. 

%

\begin{figure}[h]
\label{fig:mask}
\subfigure[Stokes I]{\label{fig:stokesi}\includegraphics[width=0.5\linewidth]{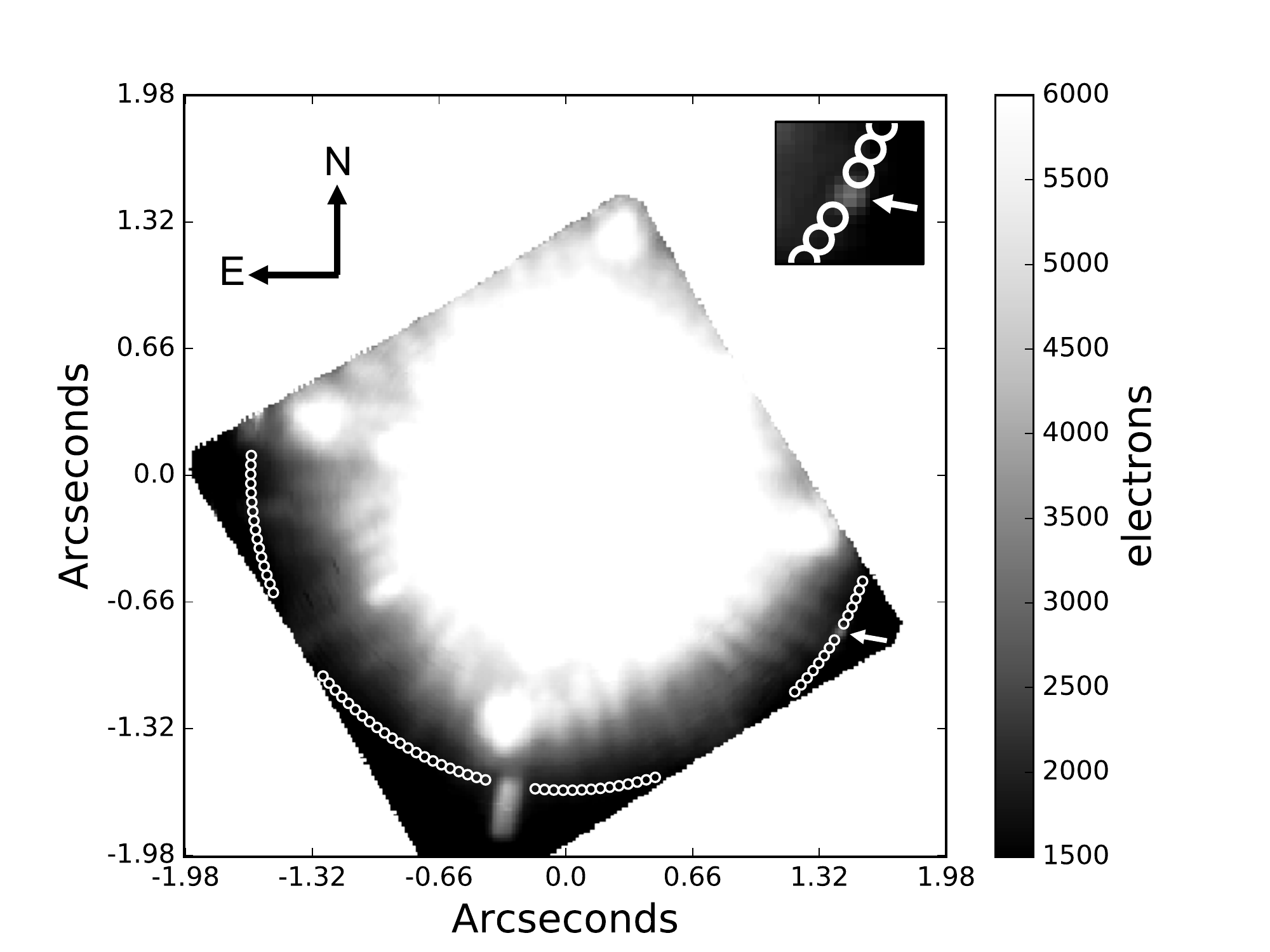}}
\subfigure[Stokes Q]{\label{fig:stokesq}\includegraphics[width=0.5\linewidth]{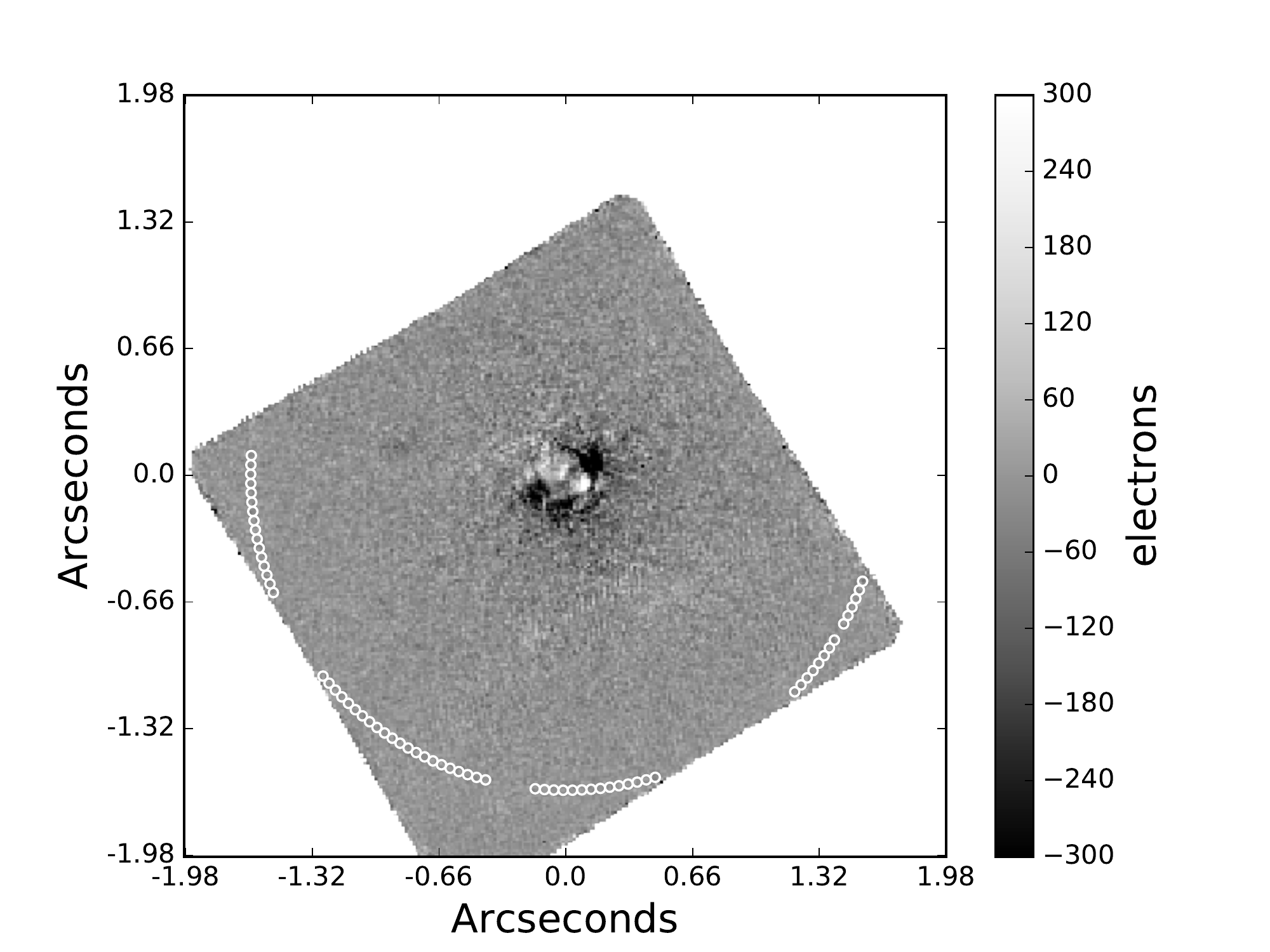}}
\subfigure[Stokes U]{\label{fig:stokesu}\includegraphics[width=0.5\linewidth]{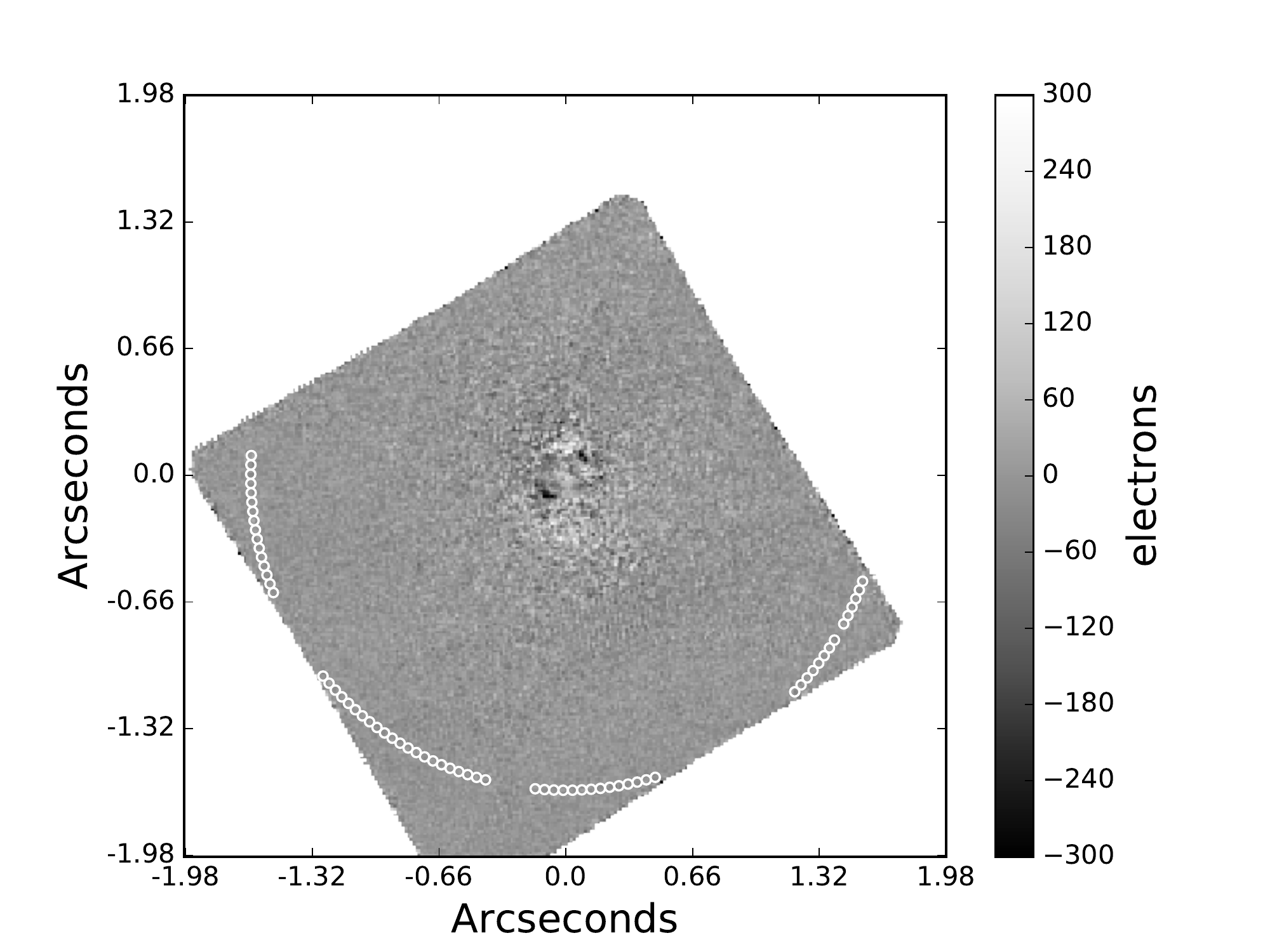}}
    \begin{minipage}[b]{0.5\textwidth}
      \caption{The reduced Stokes images with the ring of comparison apertures superimposed (the white arrow indicates the companion in Stokes I). The companion's SNR (Equation \ref{eqn:snr}) is 7.4 in Stokes I, but SNR$<1.0$ in Stokes Q and U. Hence, no polarized radiation is detected from the companion.}
      \label{fig:dummy}
    \end{minipage}    
\end{figure}

\section{Polarimetric Analysis}
\label{sec:pol}

A linearly polarized companion would produce signal in the Stokes Q and/or U frames at the same location as the companion's signal in the Stokes I frame. In order to determine whether we detect polarized radiation from HD 19467 B, we must therefore calculate the signal to noise ratio (SNR) in an aperture at the companion's location in each of the Stokes frames.

In order to choose the optimal aperture size, we first estimate the companion's full width at half maximum (FWHM) in the Stokes I frame by fitting the sum of a 2D Gaussian distribution and a plane to an $11\times11$ pixel ($\approx 3\times3$ FWHM) window around the companion's PSF. The plane component was included to account for the shape of the parent star's halo at the companion's location. The FWHM is found to be 3.44 pixels, or 0.049", which is consistent with the diffraction limit of $\lambda/$D=0.041" in the $J$ band.

The optimal aperture size for photometry will maximize the companion's signal to noise ratio in the Stokes I frame. The ``signal" is the difference of the sum of the counts inside the companion's aperture and the contribution from the parent star's residual halo at the companion's separation. This halo contribution is found by constructing a ring of apertures around the parent star at the same separation as the companion. These comparison apertures, shown in Figure 1, do not form a complete ring because the parent star was offset from the detector's center to allow for more sky rotation of the companion. We also deleted those apertures that fall within a FWHM of the bright astrometric spots. Despite these removals, the effect of small sample statistics is negligible here \citep{Mawet2014}. For each aperture size, the total number of apertures is modified to prevent them from overlapping. We then sum the flux in each of these comparison apertures. A histogram of the Stokes I aperture sums is shown in Figure 2a. The histogram is asymmetric due to the competing effects of the modified Rician statistics governing speckle noise and the whitening of those statistics after ADI reduction. We take the parent star's halo contribution to the flux inside the companion's aperture to be the mean of the comparison apertures sums, $\mu_{i}$, and the noise to be the standard deviation of the comparison aperture sums, $\sigma_{I}$. The signal to noise ratio is therefore
\begin{equation}
\label{eqn:snr}
\mbox{SNR} = \frac{I_{c} - \mu_{I}}{\sigma_{I}}
\end{equation}
where the $c$ subscript denotes the sum of the counts inside the companion's aperture ($I_{c} = 26127.4$e$^{-}$). The signal to noise reached a maximum of SNR$_I =7.4$ for an aperture diameter of $1\times$FWHM. The ``aper" aperture photometry tool provided by the IDL Astronomy User's Library was used to compute all aperture sums. 

To calculate the SNR in the Stokes Q and U frames, we use the same comparison aperture size and locations as in the Stokes I frames. The histograms of the aperture sums in the Stokes Q and U frames are shown in Figure 2b,c. We take the parent star's halo contribution to the flux inside the companion's aperture in the Q and U frames to be the means of the aperture sums, $\mu_{Q}$ and $\mu_{U}$, and the noise terms to be the standard deviations $\sigma_{Q}$ and $\sigma_{U}$ of the aperture sums. Following Equation \ref{eqn:snr}, we find that $Q_c = -99.0$ and $U_c = 41.7$, while SNR$_Q$ = 0.90 and SNR$_U$ = 0.79. We therefore conclude that we do not detect any polarized radiation from HD 19467 B. 

%

\begin{figure}[H]
\label{fig:mask}
\subfigure[$I_{c}$ - Stokes I aperture sum histogram]{\label{fig:histi}\includegraphics[width=0.5\linewidth]{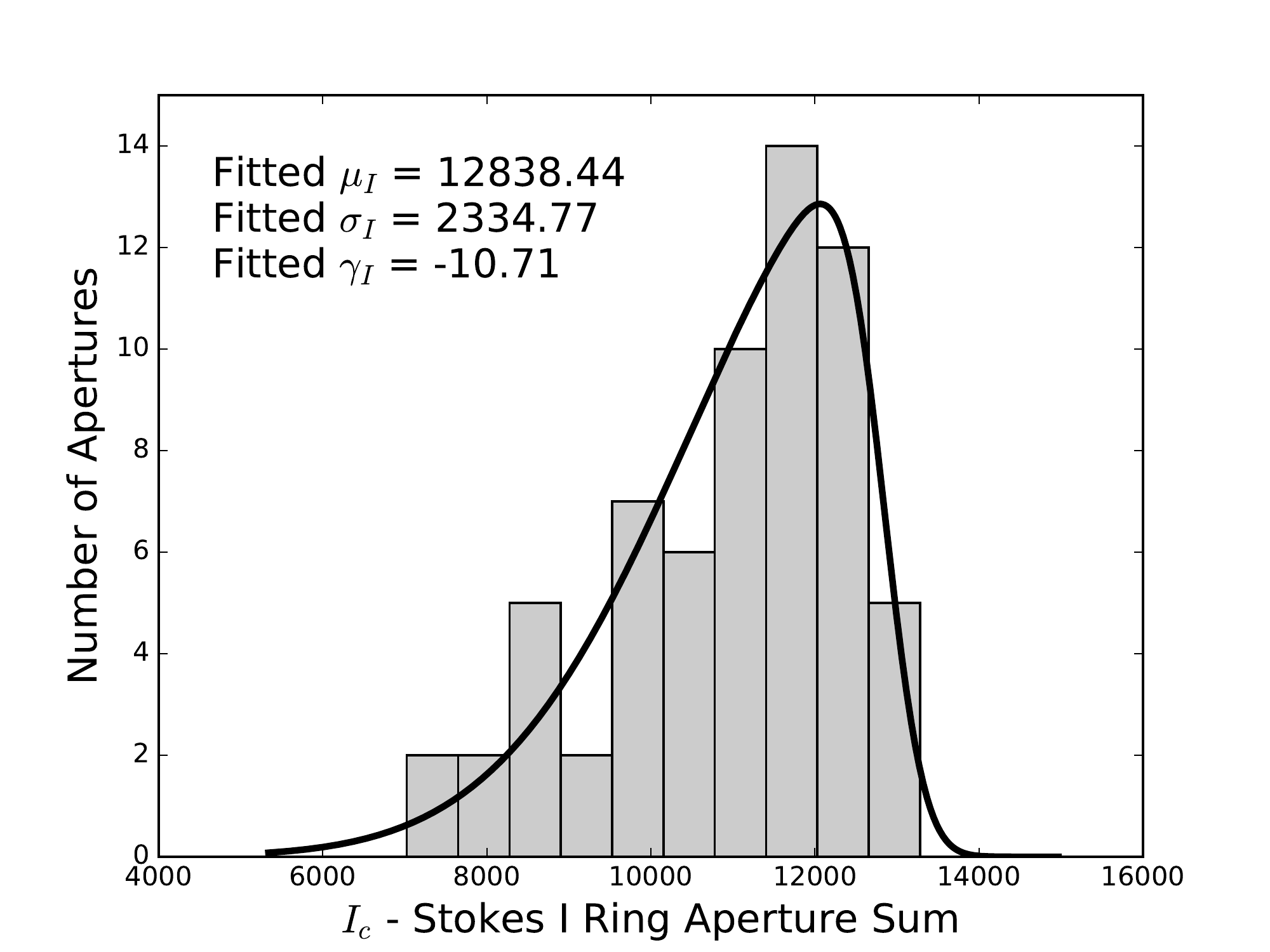}}
\subfigure[Stokes Q aperture sum histogram]{\label{fig:histq}\includegraphics[width=0.5\linewidth]{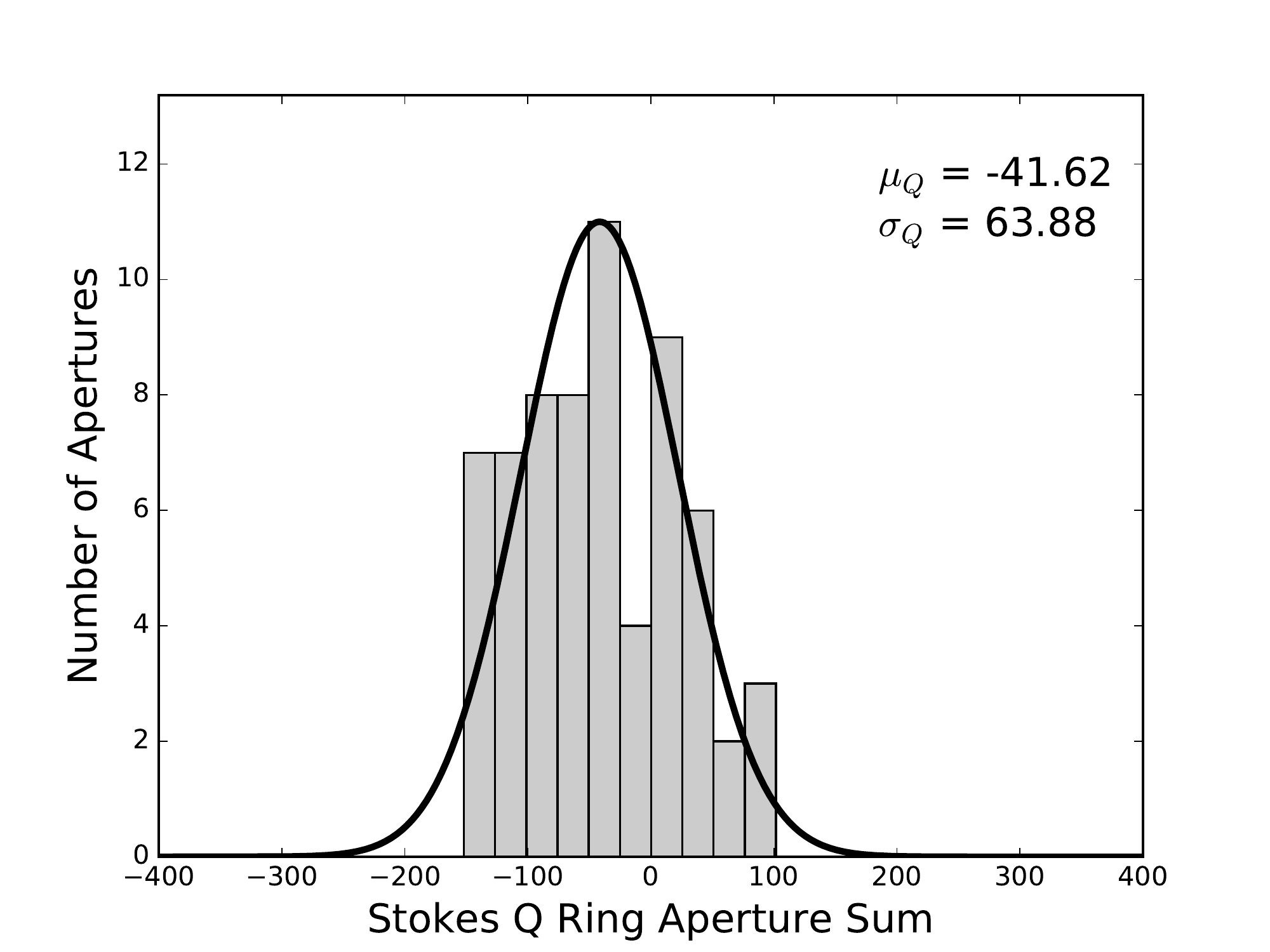}}
\subfigure[Stokes U aperture sum histogram]{\label{fig:histu}\includegraphics[width=0.5\linewidth]{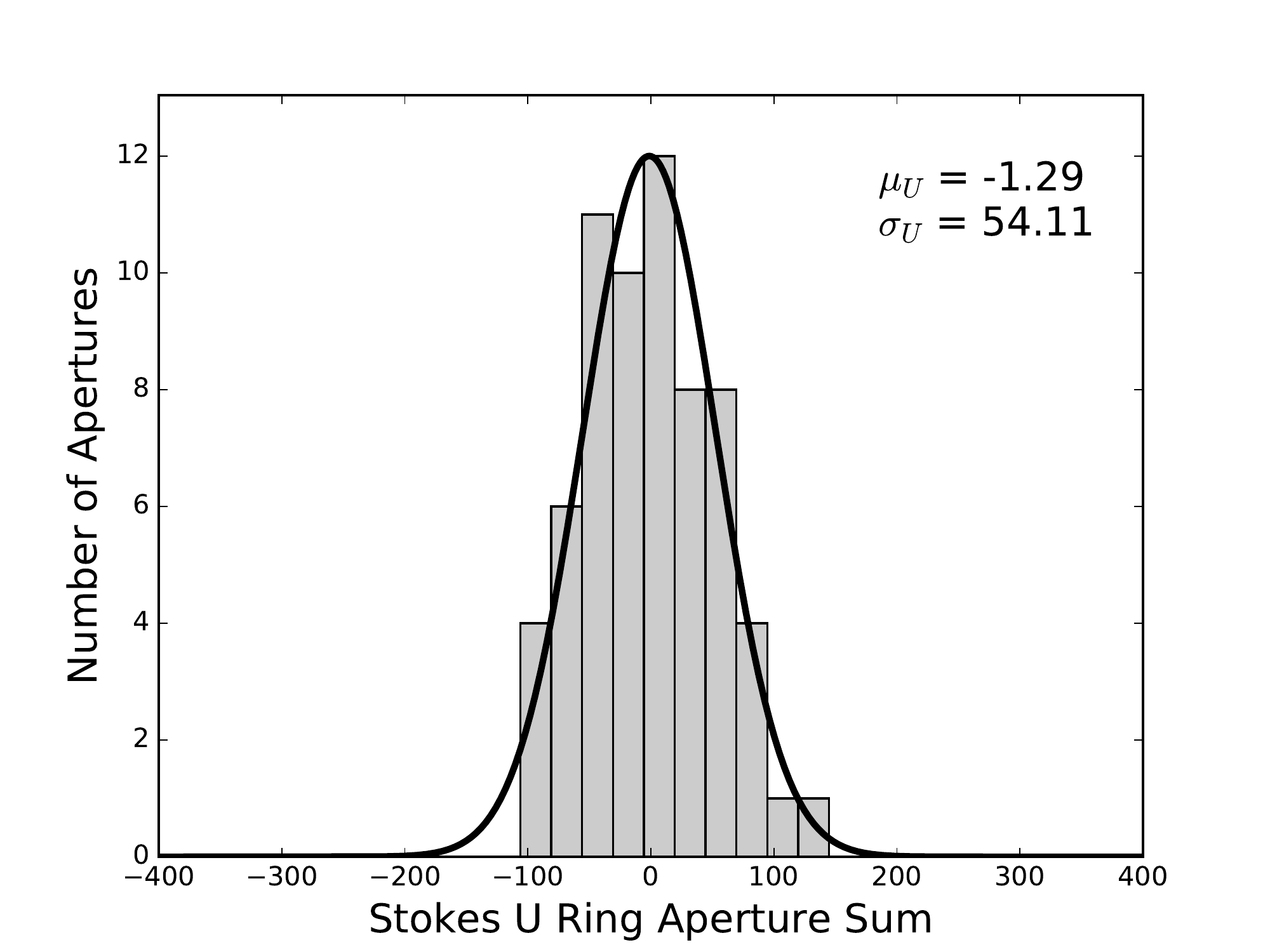}}
    \begin{minipage}[b]{0.5\textwidth}
      \caption{Histograms of the summed counts in the Stokes comparison apertures. The aperture size is equal to the full width at half maximum of the companion (3.44 pixels, or 0.049"). The Stokes I histogram (a) has an excess of higher values due to speckle noise.  Because HD 19467 A is an unpolarized star, there is little flux at the companion's separation in the Q and U frames. The large spread in Q and U values is due in part to the small number of apertures used (66).}
      \label{fig:dummy}
    \end{minipage}    
\end{figure}

Our goal now is to place an upper limit on the companion's linear polarization fraction, $p$. In general, $p$ is defined as 
\begin{equation}
p =\frac{ \sqrt{Q^2 + U^2}}{I}.
\end{equation}
Our strategy will be to compute the probability density function (PDF) of $p$, and take the representative upper limits on $p$ to be the upper bounds of the $68.27\%$, $99.73\%$, and $99.9999\%$ confidence intervals. 
 
We take the PDFs of Q and U to be Gaussian distributions with means and standard deviations matching those of the Q and U aperture sums via maximum likelihood estimation. We compute the PDF of $\sqrt{Q^2 + U^2}$ using the formalism of \citet{Aalo2007}, which considers the most general case of the PDF of $R=\sqrt{X_{c}^{2} +X_{s}^{2}}$ where $X_c$ and $X_s$ are correlated real Gaussian random variables with means $\mu_c$, $\mu_s$ and standard deviations $\sigma_c$, $\sigma_s$. \cite{Aalo2007} first rotate $(X_c, X_s)$ through the  angle $\phi$, chosen such that the new coordinates $(Y_1, Y_2)$ are uncorrelated:

\begin{subequations}
\begin{alignat}{3}
Y_1 & = & X_c \cos{\phi} + X_s \sin{\phi} \\
Y_2 & = &  -X_c \sin{\phi} + X_s \cos{\phi} \\ 
\phi & = & \frac{1}{2} \tan^{-1} \left ( \frac{2 \rho \sigma_c \sigma_s}{\sigma_{c}^{2} - \sigma_{s}^{2}} \right )
\end{alignat}
\end{subequations}
where $\rho$ is the correlation coefficient between $X_c$ and $X_s$. The full derivation of the PDF of $R$ is beyond the scope of this paper, but we quote the result here and refer to \citet{Aalo2007} for a more detailed treatment. The final PDF is given by
\begin{equation} 
\label{eqn:aalo}
\begin{split}
f_R (r) = & \frac{r}{2 \sigma_1 \sigma_2} \exp \left[-\frac{1}{2} \left(\frac{r^2}{2 \text{$\sigma_{1}$}^2}+\frac{r^2}{2 \text{$\sigma_{2}$}^2}+\frac{\text{$\mu_{1} $}^2}{\text{$\sigma_{1}$}^2}+\frac{\text{$\mu_{2} $}^2}{\text{$\sigma_{2}$}^2}\right)\right]  \\
& \sum _{n=0}^{\infty} \frac{\varepsilon_n I_n\left(\frac{r^2}{4}  \left(\frac{1}{\text{$\sigma_{2}$}^2}-\frac{1}{\text{$\sigma_{1}$}^2}\right)\right)}{\left[\left(\frac{\text{$\mu_{1} $} r}{\text{$\sigma_{1}$}^2}\right)^2+\left(\frac{\text{$\mu_{2} $} r}{\text{$\sigma_{2}$}^2}\right)^2\right]^n} \left \{ I_{2 n}\left(\sqrt{\left(\frac{\text{$\mu_{1} $} r}{\text{$\sigma_{1} $}^2}\right)^2+\left(\frac{\text{$\mu_{2} $} r}{\text{$\sigma_{2} $}^2}\right)^2}\right)\right.  \\
& \sum _{k=0}^n \delta_k C_{k}^{n} \left[\left(\frac{\text{$\mu_{1} $} r}{\text{$\sigma_{1} $}^2}\right)^2-\left(\frac{\text{$\mu_{2} $} r}{\text{$\sigma_{2} $}^2}\right)^2\right]^{n-k}  \left. \left(\frac{2 \left(\text{$\mu_{1} $} \text{$\mu_{2} $} r^2\right)}{\text{$\sigma_{1} $}^2 \text{$\sigma_{2} $}^2}\right)^k  \right \}
\end{split}
\end{equation}
where $\mu_1$, $\mu_2$ and $\sigma_1$, $\sigma_2$ are the mean and standard deviation of $Y_1$ and $Y_2$, $\varepsilon_{n=0}=1$ and $\varepsilon_{n\neq0}=2$, $\delta_{k_{\text{odd}}}=0$ and $\delta_{k_{\text{even}}}=2(-1)^{k/2}$, $C_{k}^{n}$ is the binomial coefficient, and $I_n$ is the $n$-th order modified Bessel function of the first kind. We find that Equation \ref{eqn:aalo} gives consistent results to within machine precision after five terms. 

We find that the PDF of $\sqrt{Q^2 + U^2}$ is negligibly affected by the assumption that $U$ and $Q$ are uncorrelated. The PDF is fit by a Hoyt distribution, which is equivalent to Equation \ref{eqn:aalo} for the special case of $\mu_Q = \mu_{U} = 0$.

\begin{figure}[h!]
  \centering
      \includegraphics[width=3.0in]{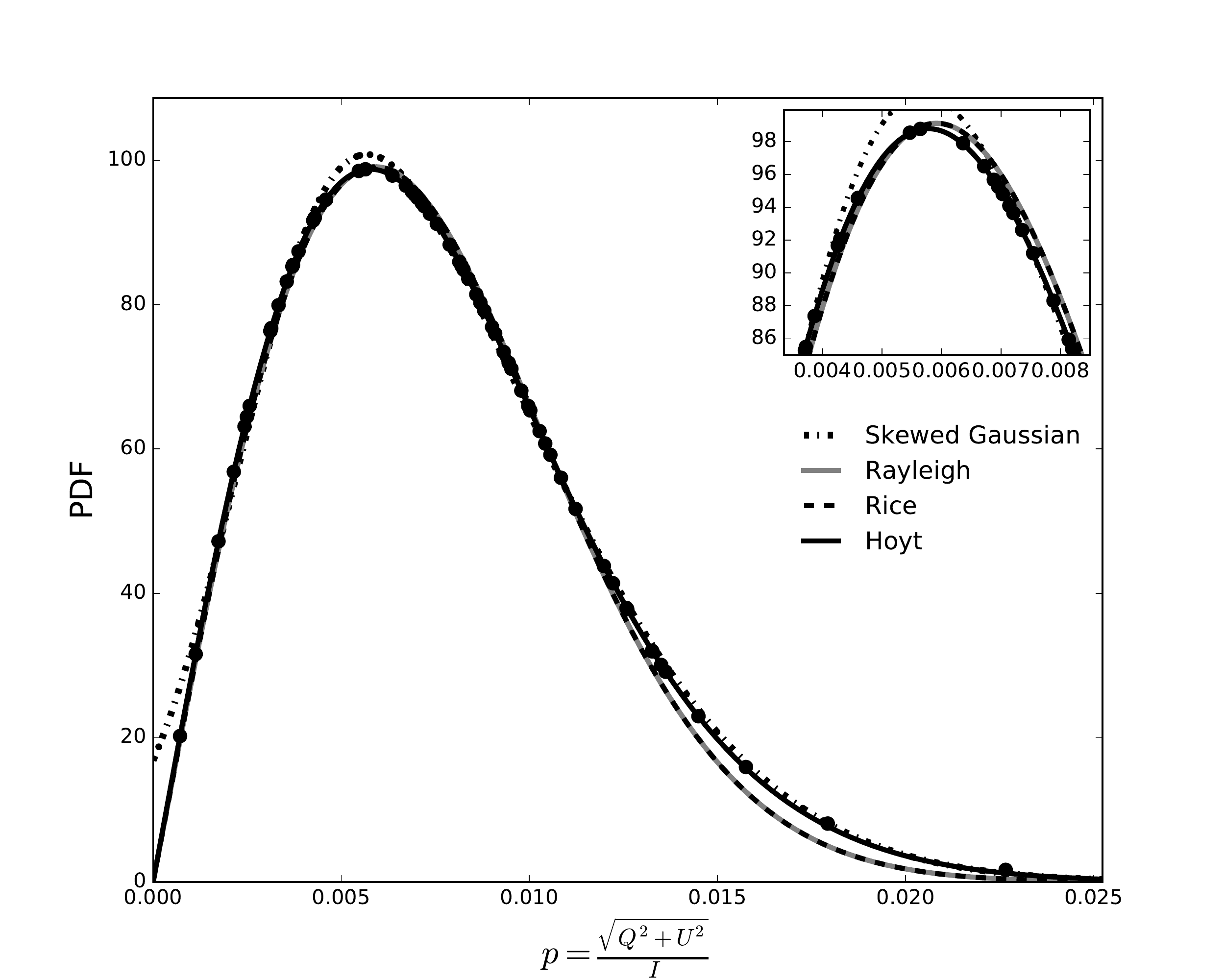}
      \caption{The probability density function of $p = \frac{\sqrt{Q^2 + U^2}}{I}$, via Equation \ref{eqn:pdist}. The four fits are to skewed Gaussian, Rayleigh, Rice, and Hoyt distributions. All but the skewed Gaussian are special cases of Equation \ref{eqn:aalo} for different values of the means and standard deviations of $Q$ and $U$. The best fit is the Hoyt distribution. }
    \label{fig:phist}
\end{figure}

To find the PDF of $p =\frac{ \sqrt{Q^2 + U^2}}{I}$, we first assume that $\sqrt{Q^2 + U^2}$ and $I$ are uncorrelated. The PDF of $I$ represents the distribution of \emph{companion} intensity values, and is therefore taken to be the counts in the companion's aperture, $I_c$, minus the distribution of counts in the comparison apertures. The comparison apertures' Stokes $I$ distribution is influenced by both the modified Rician distribution governing speckle statistics and the whitening effects of angular differential imaging. Hence, we take $p_I(i)$ to be the skewed Gaussian distribution fit to the histogram of $I_c$ - aperture $I$ sums shown in Figure 2a. Because the width, $\sigma_I$, of Figure 2a is $>10 \times$ larger than $\sqrt{I_c}$, the contribution of the companion's poisson noise to $p_I(i)$ is neglected. For R = $\sqrt{Q^2 + U^2}$, the CDF of $p=R/I$ is the probability that $R/I$ is less than a given value of $p$, or 
\begin{subequations}
\begin{alignat}{2}
F_{p}(p) = & P(R/I \leq p) \\ 
F_{p}(p) = & \int_{i=0}^{+\infty} \int_{r=-\infty}^{ip} f_{RI}(r, i) dr di + \int_{i=-\infty}^{0} \int_{r=ip}^{\infty} f_{RI}(r, i) dr di
\end{alignat}
\end{subequations}
Differentiating with respect to $p$ gives the final PDF of the linear polarization fraction:
\begin{equation}
\label{eqn:pdist}
f_{p}(p) = \int_{-\infty}^{\infty} |I| f_{R}(Ip) f_{I}(i) di.
\end{equation}
The calculated values of $f_{p}(p)$ and several fitted distributions are shown Figure \ref{fig:phist}. We find that $f_{p}(p)$ is best fit by a Hoyt distribution. Table \ref{tbl:presults} lists the upper limits to the linear polarization fraction corresponding to three representative confidence intervals calculated using the fitted Hoyt distribution. We emphasize that our analysis described in this section is distinct from that of disk polarimetry, which benefits from the assumption that light scattered by a disk is polarized in the direction perpendicular to the direction towards the central star. To our knowledge, this work presents the first analysis of point source polarimetric precision using a high contrast, integral field polarimeter. 


We now return to the question of how our limiting polarization fraction would be affected by greater sky rotation and algorithms such as PCA \citep{Soummer2012} and LOCI \citep{Lafreniere2007} that help attenuate quasi-static speckles caused by instrumental effects. The shape of the polarization fraction's PDF is governed by the PDFs of the Stokes I, Q, and U histograms shown in Figure 2. The width of the Stokes I PDF is larger than expected based on photon noise alone -- the standard deviation of the aperture values is 1479.6e$^-$, while the square root of the mean is 104.7e$^-$. This extra width is primarily due to speckle noise, which increases the distribution's standard deviation and skewness \citep{Marois2008a}. Because speckles are mostly unpolarized, however, their flux is attenuated in the Stokes Q and U frames. Visual inspection of the polarized channels in Figure 1 confirm that speckles, and hence speckle noise, are not present at the companion's separation. Indeed, \citet{Perrin2015} demonstrate that in GPI's polarimetric mode, the polarized intensity, $\sqrt{Q^2 + U^2}$, at HD19467 B's separation of 1.65" is dominated by photon noise (see also \citet{oppenheimer2008} and \citet{Hinkley2009}). Hence, greater sky rotation and PSF subtraction techniques would reduce the contribution of speckle noise to the PDF of Stokes I, but would not significantly affect the PDFs of Stokes Q and U. To assess the contribution of non-photon noise to our $p$ upper limits, we replace the PDF of Stokes $I$ values shown in Figure 2a with a poisson distribution of $\mu = \mu_{I}$. The resulting p upper limits are $0.9\%$, $1.8\%$, and $2.7\%$, for the confidence intervals of $68.27\%$, $99.73\%$, and  $99.9999\%$, respectively. Speckle reduction techniques would therefore provide some improvement to our limiting polarization fraction, but our assessment is dominated by the noise in the polarized channels.




\begin{table}[h]
\centering
\caption{Degree of linear polarization upper limits}
\begin{tabular}{ p{3.5cm} p{2.2cm}}
\hline \hline
Confidence Interval        & $p$ upper limit \\
\hline
$68.27\%$      & $1.2\%$ \\ 
$99.73\%$     & $2.4\%$ \\ 
$99.9999\%$ & $3.8\%$ \\ 
\hline
\end{tabular}
\label{tbl:presults}
\end{table}
\section{Discussion}
\label{sec:discussion}

Recent observations of photometric variability outside the L/T transition hint at the presence of cloud variability that might lead to a non-zero polarization signatures in a T5.5 dwarfs such as HD 19467 B: \citet{Radigan2014} detected $J$-band $1.6\%$ peak-to-peak variability in a red T6.5 dwarf, and $0.6\% - 0.9\%$ variability in three additional T dwarfs later than T3.5 without remarkably red colors. Photometric monitoring at $3-5\mu$m with Spitzer revealed that $19-62\%$ of T0-T8 dwarfs vary with peak-to-peak amplitudes $>0.4\%$, with T0-T-3.5 objects in the L/T transition showing no higher incidence of variability than later type T dwarfs \citep{Metchev2015}. The source of this variability may be due to variations in cloud coverage and thicknesses, or ``patchy" clouds. While it has been recently suggested that temperature variations driven by atmospheric changes below the cloud layer may also contribute to photometric variability (e.g. \citet{Robinson2014}), \citet{Apai2013} and \citet{Radigan2012} have shown that temperature fluctuations alone cannot reproduce the observed amplitudes of variability. 

The results of these photometric surveys suggest that even outside the L/T transition, T dwarfs commonly have time-varying cloud and spot weather patterns. These variations may produce polarized intensity: \citet{deKok2011} show that a fixed hotspot on the surface of an otherwise uniformly cloudy dwarf produces a higher amplitude of polarimetric variability as a function of time than flux variability. The difficulty of detecting a linear degree of polarization of less than $1.0\%$, however, has prevented observers from testing this theoretical link between photometric and polarimetric variability. For example, SIMP J013656.57+093347.3 is a T2.5 dwarf with periodic large amplitude ($50\,$mmag) variability, which is explained by invoking cool dusty cloud patches against a warmer clear background \citep{Artigau2006, Artigau2009}. However, \citet{Osorio2011} find no polarization, and place an upper limit of $p< 0.9\%$ on the $J$-band degree of linear polarization. 

In fact, only two T-dwarfs have published near-infrared polarimetric observations: the aforementioned SIMP J013656.57+093347.3 T2.5 dwarf observed by \citet{Osorio2011}, and 2MASS J12545393-0122474, a T2 dwarf observed by \citet{MilesPaez2013} with a linear degree of polarization of $p = 0.00\% \pm 0.34\%$. Our observations of HD 19467 B constitute the third polarimetric observation of a T dwarf, and the third null result. We also note that our upper limit of $p_{\text{CL}99.73\%} \leq 2.4\%$ is within a factor of a few of the previously mentioned T-dwarf upper limits, demonstrating that high contrast, integral field polarimeters are approaching the performance of direct polarimetric imaging modes. 

While these null results could be due to unfavorable viewing angles or insufficient detection limits, it is also possible that these T-dwarfs are simply cloudless, and hence would not polarize infrared radiation. Multiple scattering polarization models for cloudless T dwarfs indicate negligible polarization signals at wavelengths longer then $0.6 \mu$m at a range of inclinations and rotational velocities \citep{Sengupta2009}. These results are consistent with non-detections of near infrared T dwarf polarization. 


Clearly, more targets must be observed at higher polarimetric precisions in order to draw meaningful conclusions about the cloud properties of brown dwarfs and exoplanets. Our upper limit of $p_{\text{CL}99.73\%} \leq 2.4\%$ is within a factor of a few of the predicted $p \leq 1.0\%$ signature of a cloudy, oblate body \citep{Marley2011}, suggesting that GPI is capable of detecting polarized radiation from substellar companions given sufficient integration times to reduce photon noise in the polarized channels. To this end, our group is actively pursuing a GPI program to observe several exoplanet and brown dwarf companions at the predicted $p \leq 1.0\%$ level. Observing multiple targets will reduce the risk of non-detections resulting from unfavorable viewing angles, and may shed light on the diversity of low-mass polarimetric properties. 

\section{Conclusion}

We observed the T5.5$\pm1$ dwarf HD 19467 B in the $H$ band for 28 minutes on February 1st 2015 using the Gemini Planet Imager's polarimetry mode. We detect no polarization signal from the target, and place an upper limit on the degree of linear polarization of $p_{\text{CL}99.73\%} \leq 2.4\%$. Because this limit is larger than the predicted $p \leq 1.0\%$ signature of a cloudy, oblate body, we cannot constrain the atmospheric properties of HD 19467 B from this measurement. Our method for analyzing point source polarimetry data, however, will be applied to our upcoming GPI survey for which we expect to reach $p \leq 1.0\%$ for multiple exoplanet and brown dwarf companions. 

The future of exoplanet polarimetry is promising: because the breakup of clouds usually associated with the L/T transition occurs at lower temperatures for lower surface gravity objects, exoplanet polarimetry signals will likely benefit from the surface asymmetries introduced by patchy clouds. Indeed, \citet{Radigan2014} show that HR8799 c falls near the highest amplitude photometric variables in NIR color magnitude diagrams, a region populated by L/T transition objects. 

With the advent of modern high contrast spectro-polarimeters such as GPI and SPHERE, we are entering a new era of complementary photometric and polarimetric observations of exoplanets and brown dwarf companions. Future exoplanet polarimetric detections have the power to inform cloud particle size distributions at different pressures and corroborate the interpretations of L and T photometric survey results. Brown dwarf and exoplanet atmosphere models are currently limited by our understanding of cloud physics and its effects on observables; polarimetry is an as-yet unexploited tool to fill in these gaps. 

\acknowledgements
This material is based upon work supported by the National Science Foundation Graduate Research Fellowship under Grant No. DGE-1144469. This work was performed in part under contract with the California Institute of Technology (Caltech) funded by NASA through the Sagan Fellowship Program executed by the NASA Exoplanet Science Institute, and under the auspices of the U.S. Department of Energy by Lawrence Livermore National Laboratory under Contract DE-AC52-07NA27344

  


\end{document}